# OntologyRAG: Better and Faster Biomedical Code Mapping with Retrieval-Augmented Generation (RAG) Leveraging Ontology Knowledge Graphs and Large Language Models

Hui Feng[1][0009-0009-3486-1478], Yuntzu Yin[1][0009-0001-1996-7292], Emiliano Reynares[1][0000-0002-5109-3716], Jay Nanavati[1][0009-0003-0122-8357]

[1] IQVIA, Real World Solution, Applied AI Science, Cambridge, United Kingdom

**Abstract.** Biomedical ontologies, which comprehensively define concepts and relations for biomedical entities, are crucial for structuring and formalizing domain-specific information representations. Biomedical code mapping identifies similarity or equivalence between concepts from different ontologies. Obtaining high-quality mapping usually relies on automatic generation of unrefined mapping with ontology domain fine-tuned language models (LMs), followed by manual selections or corrections by coding experts who have extensive domain expertise and familiarity with ontology schemas. The LMs usually provide unrefined code mapping suggestions as a list of candidates without reasoning or supporting evidence, hence coding experts still need to verify each suggested candidate against ontology sources to pick the best matches. This is also a recurring task as ontology sources are updated regularly to incorporate new research findings. Consequently, the need of regular LM retraining and manual refinement make code mapping time-consuming and labour intensive.

In this work, we created OntologyRAG, an ontology-enhanced retrieval-augmented generation (RAG) method that leverages the inductive biases from ontological knowledge graphs for in-context-learning (ICL) in large language models (LLMs). Our solution grounds LLMs to knowledge graphs with unrefined mappings between ontologies and processes questions by generating an interpretable set of results that include prediction rational with mapping proximity assessment. Our solution doesn't require re-training LMs, as all ontology updates could be reflected by updating the knowledge graphs with a standard process. Evaluation results on a self-curated gold dataset show promises of using our method to enable coding experts to achieve better and faster code mapping. The code is available at https://github.com/iqvianlp/ontologyRAG.

# 1 Introduction

Biomedical ontologies represent the semantic definition and relationships between domain-specific concepts in a structural and hierarchical form [1]. Equivalent concepts are often described in different ontologies with distinctive associated relations or hierarchies to capture clinical nuances behind diverse concepts – for example, for *type-I diabetes*, a disease ontology would include the relations of its parent category such as *autoimmune disorder* and/or *diabetes*, while a drug ontology would associate it with *insulin* [2][3]. Mapping these semantically similar or equivalent concepts across ontologies – commonly recognized as code mapping – is an important step towards building a holistic formalized representation of biomedical knowledge base. [4][5][6].

It is worth noting that generating, storing and updating code mappings are extremely challenging – due to the diversity of ontology schema and source format, the demand of domain expertise to understand biomedical nuances, and the need to keep up with regular ontology updates which often contain synonym and hierarchical changes [3][7]. To achieve high quality code mapping, existing processes usually involve two main steps: using methods such as language models (LMs) specifically fine-tuned with ontology awareness to generate unrefined mappings as a list of suggested candidates, then relying on coding experts for verification and manual refinements to select the best match(es) [8][9][10]. It is very essential to have the manual refinement step as most LMs struggle to correctly map semantically similar but biomedically different concepts – for example, *acute kidney disease* can be mapped to *kidney disease*, but not vice versa. But the manual refinement is usually time-consuming and labour intensive. This is because there are usually no indicators or reasoning over mapping proximity, so even if provided with a list of top matching candidates, coding experts still need to verify each case against ontology sources and related materials [11][12]. Moreover, regular ontology updates – from ontology providers to incorporate latest research findings into ontologies – make it a demanding and recurring task to re-train LMs and re-verify selected candidates during the maintenance of database storing mapped codes.

Recently, large language models (LLMs) have shown great promises at various kinds of text-related generative tasks, with the most outstanding characteristics of these models being the general natural language understanding capability and model generalizability. However, like all models, LLMs struggle at external or unseen knowledge as well as out-of-date internal knowledge. Moreover, regular re-training or fine-tuning LLMs to adapt to ontology source changes would be costly and labour demanding – because it requires frequent training data curation, expensive computational set-up, and experts for model handling.[13] On the other hand, knowledge graphs (KGs) have shown effectiveness in storing rich semantic and hierarchical information from ontologies, it is also straightforward to update information stored in KGs.[14][15] However, accessing stored information requires knowledge of query language and specific graph metadata, meanwhile the retrieved graphs are often hard to interpret. Studies have shown that LLMs and KGs could complement with each other's limitations when used together, such as in a retrieval-augmented generation (RAG) system, where LLMs can be used for retrieving, reasoning over, and lexicalising the information encoded in knowledge graphs, to enhance the accessibility and interpretability of graphs [16][17][18], to offer reliable and up-to-date external knowledge, thus providing great excellence and convenience in numerous applications.[19][20]

In this work, we propose a customised ontology-enhanced retrieval-augmented generation pipeline (OntologyRAG) that leverages the in-context reasoning capabilities of off-the-shelf LLMs by infusing ontology knowledge graphs to enable coding experts executing better and faster code mapping. This pipeline contains three parts: indexing (converting ontology source files into standard format, generating unrefined mappings, and storing the information as KGs in a database), retrieving (generating SPARQL queries to retrieve subgraphs from the database) and reasoning on retrieved results (providing mapping proximity and summary). Through evaluation against an expert curated gold dataset, we show promises to significantly enhance the quality and efficiency of code mapping with OntologyRAG, by jointly exploiting the structured information encoded in ontology knowledge graphs and the language generation capabilities of LLMs.

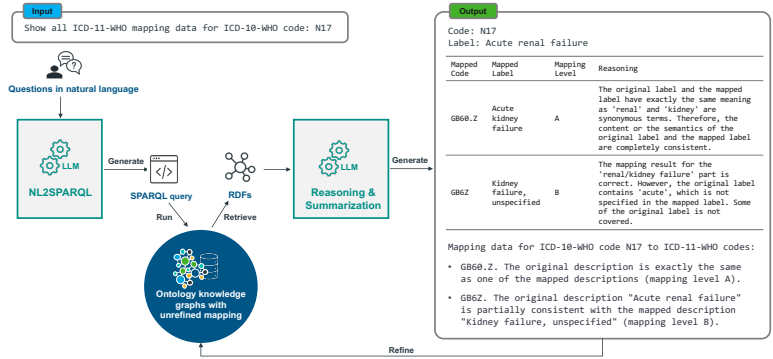

**Fig. 1.** Illustration of our proposed Ontology-Enhanced Retrieval-Augmented Generation (OntologyRAG) workflow for code mapping.

## 2 Methods

Our proposed approach leverages LLMs at its core. In our work, we first carried out an ablation study on assessing the direct code mapping prediction capability of LLMs, then moved onto the construction and evaluation of the RAG pipeline. The final OntologyRAG prototype stores ontology information such as code descriptions, relations and unrefined mappings between ontologies into RDF knowledge graphs during the indexing process, and retrieves these information with an LLM-enabled method that allows users to query the database with natural language questions (NLQs). The response from the prototype is a text summary containing retrieved results, code mapping levels and associated reasonings (**Fig. 1**).

### 2.1 Ablation Study on LLMs for Code Mapping without KGs

To understand the out-of-the-box capabilities of LLM at executing the code mapping tasks, we curated a gold dataset containing 500 mappings from ICD-9-CM to the 2018 version of ICD-10-CM. An ablation study was carried out by running zero-shot experiments with the selected LLMs at default temperature, with the task instructions in the prompt being *"Task Summary: You are a clinical coder and are assigning ICD10CM codes to existing ICD9CM codes. Instructions: Please assign the corresponding ICD10CM code to the ICD9CM code provided based on the 2018 version. If multiple ICD10CM codes can be mapped, please list them all"*.

We consider the prediction correct if the response from LLM contains the correct mapped code. If multiple ICD-10-CM codes are listed in the gold dataset, we credit the model for each correct prediction and do not penalize it for returning extra codes. For example, if the gold dataset annotated codes are *[code_A, code_B]*, and the LLM predicted codes are *[code_A, code_C, code_D]*, we consider the prediction of *code_A* being correct and *code_B* being incorrect, thus the accuracy for this prediction to be 50%.

Two OpenAI models (GPT-3.5-Turbo and GPT-4) [21] and two open-source models (Meta-Llama-3-8B [22] and Google-Flan-T5-XXL [23]), which have shown to perform well on Biomedical Language Understanding and Reasoning Benchmark (BLURB) tasks [24], have been evaluated. Each experiment was repeated three times, and the mean average accuracy from all data points was used as evaluation metrics.

## 2.2    Ontology Knowledge Graph Generation and Indexing

For each ontology of interest, the prerequisite to use the pipeline is to construct a knowledge graph that captures the hierarchical structured information (features, relations, and unrefined mappings) of the new and existing ontologies. We build such a knowledge graph by applying the ETL (Extract, Transform, Load) processing schema.

At the *extract* stage, we obtain the data from the original format by performing various tasks depending on the source – from unzipping a file to executing dedicated SQL scripts. For instance, we retrieved the International Classification of Diseases, Clinical Modification, Ninth and Tenth Revisions (ICD-9-CM and ICD-10-CM).[1] The ICD-9-CM diagnosis and procedure codes and labels were retrieved from the Centers for Medicare and Medicaid Services (CMS) official website [25]. The ICD-9-CM files present the codes and labels of the diagnoses and procedures in a two-column table. The ICD-10-CM diagnosis and procedure codes and labels were retrieved from the Centers for Disease Control and Prevention (CDC) official website [26]. The ICD-10-CM files present the codes and labels of the diagnoses and procedures in XML files under a proprietary schema. The ICD-9-CM to ICD-10-CM mappings - known as General Equivalence Mappings (GEM) files - were retrieved from the CMS official website [27]. The GEM files are typically provided in plain text format, with each line representing a mapping entry. Each entry includes the source code, target codes, and any relevant flags or attributes, e.g., one-to-one, one-to-many, or many-to-one relationships, and flags to indicate whether a mapping is approximate or exact.

---

[1] ICD-9-CM and ICD-10-CM are based on the corresponding versions of the World Health Organization International Classification of Diseases (ICD-9 and ICD-10). ICD-9-CM was the official system for assigning codes to diagnoses and procedures associated with hospital utilization in the United States. It was used to code and classify mortality data from death certificates until 1999, when the use of ICD-10 for mortality coding started.

Then, a customized-per-source *transformer* is executed to generate an RDF-based representation from the extracted data. RDF is a standard model that provides a structured way to describe and interlink data, developed by the World Wide Web Consortium (W3C) [28]. RDF data is organized as triples, each consisting of a subject, predicate, and object, forming a directed graph where nodes represent resources and edges represent relationships between them. The triple structure allows for a flexible and extensible way to represent information, making RDF particularly useful in use cases like ours where data from various sources need to be integrated and used together. At the time of writing this work, we had not completed the established internal process to open source the code base of the transformer stage, although the output of such a process can be found on the mentioned GitHub repository.

Finally, we *load* the RDF data into Oxigraph [29]. Oxigraph is an open-source graph database written in Rust and based on the RocksDB key-value store [30]. It provides a set of utility functions for reading, writing, and processing RDF files. This RDF data can be manipulated and retrieved through queries written in the SPARQL Protocol and RDF Query Language (SPARQL) [31]. SPARQL is a standard developed by the W3C, whose queries have a syntax like SQL and are based on matching triple patterns against RDF data. SPARQL supports four main types of queries: SELECT, CONSTRUCT, ASK, and DESCRIBE. The SELECT query retrieves specific data elements and returns a table of results, like SQL. The CONSTRUCT query retrieves data elements and generates a new RDF graph from those elements. The ASK query checks if a specific pattern exists in the data and returns a boolean value, indicating whether the pattern was found. Lastly, the DESCRIBE query retrieves an RDF graph that describes a specific resource or group of resources, providing detailed information about them.

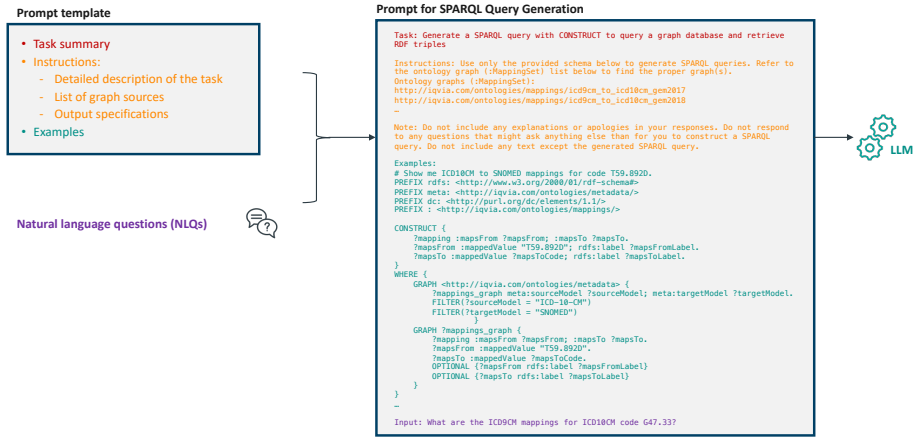

**Fig. 2.** Prompts for generating SPARQL queries in the NLP2SPARQL module, including the input NLQ, graph sources and NLQ-to-SPARQL examples.

### 2.3  Subgraph Retrieval

Once the ontology graphs are ready, as mentioned in 2.2, SPARQL queries are required to retrieve subgraphs relevant to ontology related questions from the database.

However, it requires both professional SPARQL query construction skills and familiarity to the knowledge graphs to construct valid and correct SPARQL queries for result retrieval. To remove this technical bar, we create the NL2SPARQL module in the OntologyRAG pipeline, which takes in an NLQ and returns a SPARQL query, so that NLQ could be directly used to retrieve information from the ontology graph database.

The NL2SPARQL module takes an NLQ as input, which is then added as the *Input* parameter into a prompt template detailing task instructions, a list of graph sources, and a few NLQ-to-SPARQL examples to create a real-time prompt for LLM to generate a SPARQL query (**Fig. 2**). The SPARQL generation process also includes a self-validation checkpoint to check the validity of the generated SPARQL. In addition to verifying whether the output has the correct SPARQL syntax, it also verifies whether the graph source, entities, and relationships described by the query exist in the ontology graph database. If the generated SPARQL query is invalid, it will automatically repeat the SPARQL generation process until the query is valid and can be applied to retrieve information from the graph database.

### 2.4 Mapping Proximity Assessment and Summary Generation

As mentioned in section 2.2, the indexed mappings between ontologies in the KG database are unrefined mappings, which means there might be incorrect mappings between codes and coding experts will be needed to validate and update these mappings for high quality refined mappings that could be used for other biomedical or clinical tasks.

To assist the validation process, we create the Reasoning & Summarization module which takes the NLQ and code pairs (queried code in NLQ and retrieved code by NL2SPARQL) as input, assigns one of the three mapping levels in **Table 1** based on semantic similarity and logical reasoning, then returns a natural language summary as output (**Fig. 3**).

Table 1. Definition of mapping levels for code pairs.

| Mapping level | Definition | Example[2] |
|---|---|---|
| A | The content or the semantics of the original label and the mapped label are completely consistent. | Q.: acute renal failure R.: acute kidney failure |
| B | Parts of the original and the mapped labels are related, but it is not certain whether they match or conflict. | Q.: renal failure R.: acute kidney failure |
| C | The original and the mapped labels partially conflict with each other. | Q.: acute renal failure R.: chronic kidney disease |

In the reasoning step, we made the mapping level prediction task independent of reasoning to enhance the accuracy of LLM's prediction performance for ontology pairs. LLM is prompted twice to 1) predict the mapping level between the two disease descriptions of the retrieved code pairs; 2) give reasoning based on the mapping level predicted above (**Fig. 3**). Finally, the input NLQ, retrieved code pairs, predicted mapping levels as well as reasoning are fed to LLM one last time, summarizing all code mapping results.

---

[2] Q.: queried code. R.: retrieved code.

To compare the mapping-level prediction capabilities of different models in the Reasoning & Summarization module, a gold dataset containing 500 pairs of disease descriptions and their corresponding mapping levels was created. When creating the gold dataset, we first used GPT-3.5-Turbo to generate 500 disease description pairs based on common disease classification systems and predict a mapping level to the disease description pairs based on semantic similarity as described in **Table 1.**, then performed manual revision by domain experts. Each record in the gold dataset contains two disease descriptions and a mapping level describing the semantic proximity.

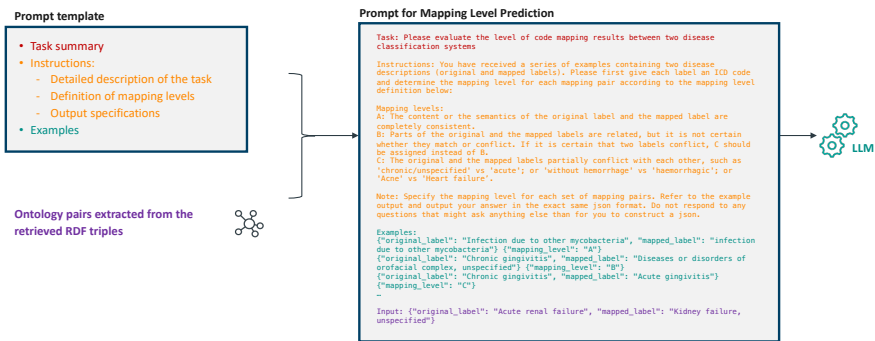

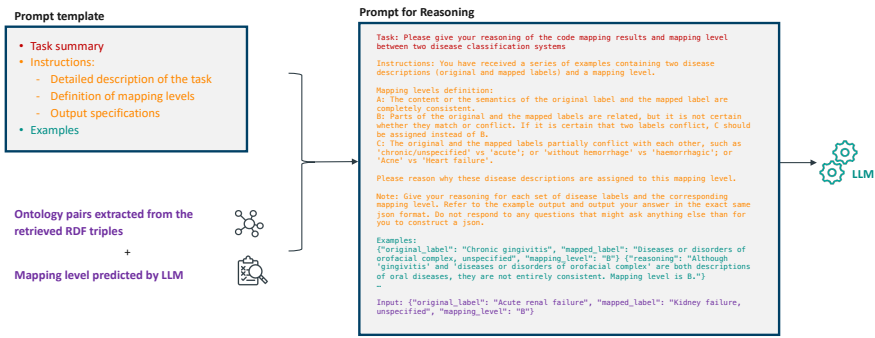

**Fig. 3.** Prompts for 1) mapping level prediction and 2) reasoning in the reasoning step of the Reasoning & Summarization module.

Four distinct prompting strategies were compared for all the models mentioned in section 2.1: 1) zero-shot: no examples; 2) few-shot: 16 examples; 3) example enhanced few-shot: 21 examples, with 5 more examples for mapping level B, as B represents partially semantic similarity or uncertain cases, which covers more diverse combinations than A or C; 4) chain-of-thought with example enhanced few-shot: based on example enhanced few-shot (21 examples). Each experiment was repeated three times, and the mean average accuracy and per mapping level precision from different LLMs and prompting strategies were used as evaluation metrics.

Gold datasets used for evaluation, prompt templates, scripts to run the retrieval and reasoning pipeline, and video demonstration of the pipeline can be accessed at https://github.com/iqvianlp/ontologyRAG.

## 3      Results and Discussions

### 3.1    Ablation Study on LLMs for Code Mapping without KGs

As shown in **Table 2**, for the ablation study with our gold dataset, when used zero-shot prompting asking LLMs to directly return mapped codes, none of the selected models achieved more than 10% overall accuracy, with the best performing model (GPT-4) achieving only 8.38%.

Besides LLMs' the inherent limitation of hallucinations, this poor performance is probably also related to the fact that most ontology source files are proprietary data – meaning most LLMs might not have been trained on such data – as well as the domain-specific semantic complexity encoded behind every ontology code.
Our ablation study not only shows the complexity of ontology code mapping tasks but also reveals the huge gap of using only LLMs for such tasks.

Table 2. Accuracy (%) from using different LLMs directly for code mapping

| Prompting strategy | GPT-3.5-Turbo | GPT-4 | Meta-Llama-3-8B-Instruct | Google-Flan-T5-XXL |
|---|---|---|---|---|
| Zero-shot | 0.27±0.21 | 8.38±1.09 | 4.67±1.70 | 0.00±0.00 |

### 3.2    SPARQL Query Generation and Subgraph Retrieval

For the NL2SPARQL module, we experimented with three SOTA instruction-tuned LLMs: GPT-3.5-Turbo, GPT-4, Meta-Llama-3-8B. The LLMs were used without further fine-tuning, and the results indicated that all three models were able to generate accurate SPARQL queries to retrieve information from correct knowledge graphs within two attempts.

However, there were slight differences in the format of the responses returned by the LLMs. While the prompt instructed the models to solely output the generated SPARQL query without any irrelevant text, only GPT-3.5-Turbo and GPT-4 were able to provide a specified response as requested, with an out-of-the-box, ready-to-use SPARQL query without any further post-processing. In contrast, Meta-Llama-3-8B tended to include additional dialogue or replies to the prompts in its output, which required some light post-engineering work to extract the SPARQL query embedded in the response.

### 3.3 Evaluation on Mapping Proximity Assessment and Reasoning

For the Reasoning & Summarization module, we collected and curated a gold dataset of 500 records for performance evaluation, with each data point containing two disease descriptions and a manually assigned mapping level. The impact of model choice, prompting strategy and temperature on the overall accuracy of mapping level prediction against the gold dataset was evaluated. We found few-shot learning help improve the performance of all tested models when compared with zero-shot learning. Although CoT prompting yielded better performance for both GPT models and Meta-Llama-3-8B-Instruct, it yielded worse performance for Google-Flan-T5-XXL. This might be contributed by Google-Flan-T5-XXL being the only decoder-encoder model while the other three being decoder-only models. We also noticed the performance of Meta-Llama-3-8B-Instruct with both few-shot prompting strategies exceeded GPT-3.5-Turbo and Google-Flan-T5-XXL and was close to GPT-4 (**Table 3**), which is impressive given the number of model's parameters, making it a promising open-source candidate to execute this task.

**Table 3.** Accuracy (%) for code mapping level predictions with different prompting strategies[3]

| Prompting strategy | T.* | GPT-3.5-Turbo | GPT-4 | Meta-Llama-3-8B-Instruct | Google-Flan-T5-XXL |
|---|---|---|---|---|---|
| Zero-shot | 0.2 | 71.47±1.62 | **81.07±0.61** | 58.67±2.20 | 63.80±1.22 |
|  | 0.6 | 72.73±0.81 | **81.47±0.90** | 59.80±0.00 | 63.73±2.25 |
|  | 1 | 72.47±1.10 | **81.07±0.23** | 58.67±2.20 | 61.53±1.50 |
| Few-shot | 0.2 | 72.47±1.33 | **86.33±0.12** | 82.20±0.72 | 63.40±0.40 |
|  | 0.6 | 74.47±1.29 | **85.80±0.87** | 80.47±0.64 | 60.13±1.22 |
|  | 1 | 73.93±0.92 | **86.20±0.20** | 82.00±0.00 | 63.20±0.00 |
| Example enhanced few-shot (Enhanced) | 0.2 | 79.20±1.78 | **85.40±0.80** | 80.47±0.64 | 64.00±0.92 |
|  | 0.6 | 79.73±2.55 | **86.07±1.33** | 79.33±0.70 | 61.80±0.92 |
|  | 1 | 78.13±0.46 | **85.53±0.58** | 80.00±0.00 | 63.40±0.00 |
| Chain-of-thought (CoT) | 0.2 | 79.13±0.50 | **87.13±1.01** | 76.13±0.42 | 55.47±1.30 |
|  | 0.6 | 79.07±0.50 | **86.13±0.81** | 76.33±0.23 | 53.00±1.51 |
|  | 1 | 77.93±2.32 | **87.47±0.83** | 75.80±0.00 | 56.80±0.00 |

Since these mapping levels are set to help coding experts focus on ambiguous or complex cases, we needed to consider the impact of each mapping level for human reviewers. For example, high precision prediction on levels A and C indicates the possibility of introducing an automatic generation of group filters to allow reviewers directly accepting or rejecting these more semantically obvious mapping results. This could enable these coding experts to focus only on those mapping results classified as level B, which are the difficult and ambiguous cases, thus accelerate the code mapping process.

---

[3] T.*: Temperature.

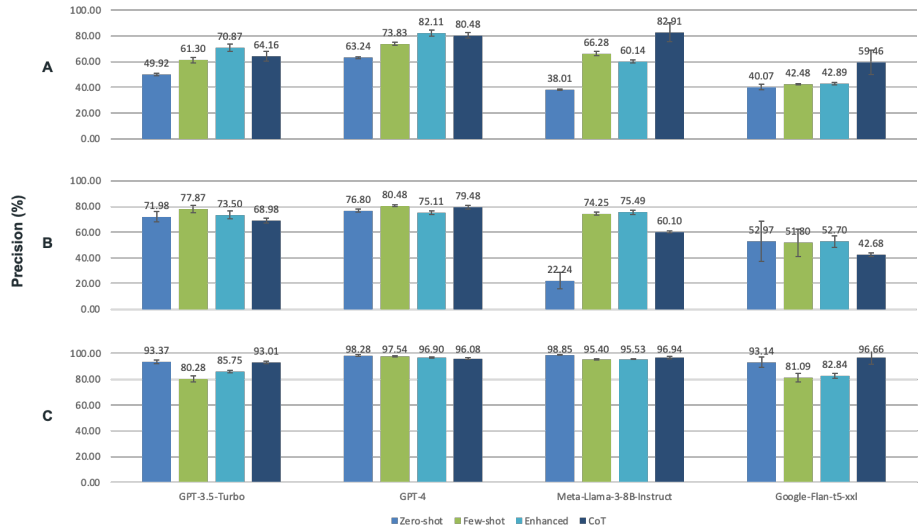

**Fig. 4.** Precision (%) of biomedical code mapping at each level with different prompting methodologies.

Given the need to weigh the three mapping levels differently, the overall accuracy of the LLMs should not be the only metrics for evaluating their mapping level assignment performance. To help gauge a better suited model for this module, we incorporated the precision of each mapping level prediction as another evaluation metrics and prioritized the models that can provide higher precision for levels A and C.

All tested combinations of LLMs and prompting methods achieved a relatively high precision (>80%) for level C prediction. Outstandingly, all combinations for GPT-4 and Meta-Llama-3-8B achieved > 95% precision. The chain-of-thought (CoT) prompting strategy - asking LLMs to perform reasoning before making level predictions – yielded the highest prediction precision from both open-source LLMs for mapping level A, with Meta-Llama-3-8B even surpassing GPT-4 with a score of nearly 83% (**Fig. 4**).

Distinguishing levels A and B appeared to be more challenging for the LLMs. This could be seen as the models view the two codes as closely related and consistent but still have difficulty determining if they are an exact match. We took a deeper look at the results from these models with a confusion matrix to help better understand the misclassification details (**Fig. 5**). From the matrix, we could see the impact from prompting engineering on predicting mapping level C is much less than the other two levels. We could also see that, with zero-shot approach, Meta-Llama-3-8B and Google-Flan-T5-XXL tend to predict mapping level A and C over mapping level B, meanwhile misclassify many mapping level B cases to mapping level A, while GPT-3.5-Turbo and GPT-4 have a better prediction balance amongst all three mapping levels. However, promisingly, few-shot, few-shot enhanced and CoT approaches all effectively helped Meta-Llama-3-8B "comprehend" the nuances in differentiating mapping level A and mapping level B, thus significantly improved its prediction precision for these two levels. These prompt engineering strategies also further enhanced GPT-3.5-Turbo and GPT-4's prediction precision on mapping level A and B. However, such improvements are not observed with Google-Flan-T5-XXL.

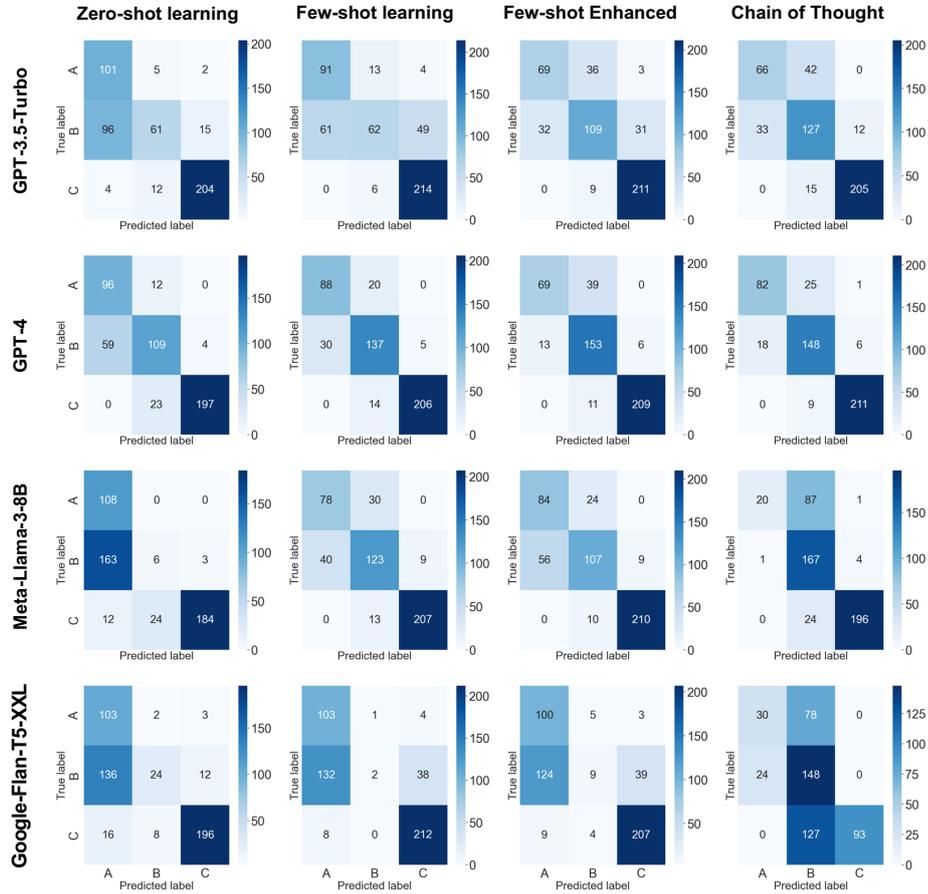

**Fig. 5.** Confusion matrix of biomedical code mapping level predictions using different prompting strategies (temperature=0.2).

As mentioned in section 2.4, to help coding experts understanding the rationale behind the predicted mapping level, all LLMs were also prompted to give a natural language reasoning to support the prediction. To evaluate the generated reasoning output, we have considered using automated evaluation with metrics such as rouge-L. However, due to the generative nature of LLMs, it was extremely difficult to predict how the natural language responses could be constructed. Therefore, it was very difficult to generate a fair gold standard with expected reasoning answers to evaluate the quality of the natural language reasoning output automatically. Although we believe having a robust method to construct a gold dataset with automated evaluation to assess reasoning output could be an interesting piece of work, it is out of scope for this study but might be of our future interest. For this study, we manually examined the reasoning responses and found the general quality of the reasoning very closely associated with mapping level prediction. Therefore, we believe presenting the result of mapping level is sufficient to show the effectiveness of different LLMs in executing the tasks associated with the Reasoning & Summarization module.

## 4    Conclusion

In this work, we created OntologyRAG, a pipeline that leverages ontology knowledge graphs and the in-context-learning capabilities of LLMs to enable coding experts achieving better and faster biomedical code mapping. The key contributions of our work are:

- Created OntologyRAG, an LLM-based RAG pipeline for ontology code mapping. Demonstrated its effectiveness over indexing and retrieving complex ontology related information and showed its promises to assist code mapping refinement.
- Provided two gold datasets, one for evaluating the accuracy of model's capability at direct code mapping, and the other for evaluating the effectiveness of model's capability at code mapping proximity prediction.
- Evaluated the effectiveness of several state-of-the-art LLMs against our gold datasets and provided the findings.

The OntologyRAG pipeline is scalable and generalizable. It contains a ready-to-use index module that can convert and store most ontology source files and mapping files into knowledge graph database; a retrieval module that takes natural language questions as input and automatically generates valid SPARQL queries to retrieve question-related information from the graph database; and a reasoning and summarization module that assesses the relevance of the retrieved information to the input question then output the relevance proximity with natural language reasoning. The choice of LLMs in our pipeline is flexible. The gold datasets and evaluation methods we provided could help decide if upgrading LLMs was needed for during pipeline maintenance.

Our results show great promises in incorporating off-the-shelf LLMs such as Meta-Llama-3-8B and GPT-4 to enhance the accessibility and interpretability of complex information encoded in ontology knowledge graphs. By accepting natural language questions and providing output containing retrieved results and pre-analysed mapping levels and reasoning, the pipeline allows coding experts to concentrate their effort on the most ambiguous and complex mapping cases instead all cases, thereby enhancing the mapping efficiency and quality.

This research prototype demonstrates the promising integration of LLMs with KGs to assist code mapping refinement. The pipeline facilitates rapid and flexible adaptation of future LLMs with enhanced performance and ensures swift responsiveness in updating knowledge graphs to reflect ontology changes. Future research will focus on exploring how to extend and enhance the prototype into a production level application to facilitate fast and robust ontology mapping.